\documentclass[
prl,
twocolumn,
aps,
]{revtex4}
\usepackage{amsmath,amssymb,amsfonts}
\usepackage{graphicx}

\begin{document}
\title{Complexity perspectives: an anomalous diffusion approach}
\author{Guilherme R. Rezende$^{1,2}$}
\email[E-mail: ]{guirrezende@gmail.com}
\author{Luciano C. Lapas$^{1,3}$}
\email[E-mail: ]{luciano@fis.unb.br}
\author{Fernando A. Oliveira$^{1}$}
\email[E-mail: ]{fao@fis.unb.br}

\affiliation{
$^{1}$Instituto de F\'{\i}sica and Centro Internacional de F\'{\i}sica da Mat\'{e}ria Condensada, Universidade de Bras\'{\i}lia, Caixa Postal 04513, 70919-970 Bras\'{\i}lia, Distrito Federal, Brazil.\\
$^{2}$ Escola T\'{e}cnica Federal de Bras\'{\i}lia, 70730-521 Bras\'{\i}lia, Distrito Federal, Brazil.
$^{3}$ Departamento de Matem\'{a}tica, Universidade de Bras\'{\i}lia, 70910-900 Bras\'{\i}lia, Distrito Federal, Brazil.
}

\begin{abstract}
The science of complexity is far from being fully understood and even its foundations are not well established. On the other hand, during the last decade, the random motion of particles or waves -- the so-called diffusion -- has been known better. In this paper, we discuss how simple ideas of diffusion can be used to deal with the description of most complex structure.
\end{abstract}


\maketitle

\section{Introduction}
Complexity is still in its begin due to the large class of proposal and problems it reaches. Some few attempts have been done in order to measure complexity~\cite{Pincus91,Rosa99}. However, complexity is ``build up'' from disciplines such as non-equilibrium statistical mechanics, non-linear dynamics, and sciences as diverse as geophysics and biology, which are ``complex'' themselves, and consequently any progress in its branches will contribute to its conceptual evolution. Non-equilibrium statistical mechanics is somehow the simplest part of this high block, and its evolution shines light on all those connected area. Most developed part of the non-equilibrium statistical physics is concerned with relaxation process, \emph{i.e.} the way a given system relax to equilibrium or to a steady state. Few theorems have been established for that, mainly from linear response theory were one can compute the ansatz of a system to a small perturbation~\cite{Kubo91}.

Complex systems are usually nonlinear, inhomogeneous, and far from equilibrium. A common feature of those systems are pattern formations~\cite{Cross93}, where the translation symmetry is broken. Pattern formations are common as well in diffusion of alive beings~\cite{Fuentes03,Giuggioli03}, and it has been demonstrated~\cite{daCunha09a,daCunha09b} that the violation of ergodicity presents a situation similar to that of anomalous diffusion~\cite{Costa03,Lapas07}. This is a green signal to extend those ideas to more complex systems. However, we need be careful to not drive ourselves in wonderful, but unrealistic generalizations. Ergodic hypothesis (EH) is a fundamental issue in statistical physics~\cite{Boltzmann74,Khinchin49,Lee01,Lee07a} and it validates a large number of theorems in physics. As a consequence it becomes a branch of mathematical physics, and a large number of works dealing with EH have been done by mathematicians, with small value for those interested in the analysis of mensurable quantities. In this context the Khinchin theorem (KT) is of special importance, since it was formulated in terms of correlations functions which can be obtained by use of linear response theory, and they are also directly connected with experiments, such as light or neutron scattering~\cite{Oliveira81}.

\section{Anomalous diffusion and complexity}
Diffusion is of fundamental importance for physics and much more simple to describe than reaction rates~\cite{Kramers40,Hanggi90,Oliveira98}, besides being the focus of extensive research in many interdisciplinary sciences~\cite{Astumian02,Morgado02,Bulashenko02,Costa03,Bao03,Budini04,Dorea06,Bao06}. An usual manner to study the diffusive dynamics is to deal with the mean square displacement of the particles, given by
\begin{equation}
\lim_{t \rightarrow \infty} \left\langle x^{2}\left( t\right) \right\rangle \propto t^{\alpha }\text{,} \label{eq:x2}
\end{equation}
where $\left\langle \ldots \right\rangle$ denotes an ensemble average. The exponent $\alpha$ classifies the type of diffusion: for $\alpha =1$, we have normal diffusion; for $0<\alpha <1$, subdiffusion; and $\alpha >1$, superdiffusion. There is another way to deal with diffusion by considering a Brownian Motion described by a generalized Langevin equation (GLE), or Mori formalism~\cite{Mori65}, of the form
\begin{equation}
\frac{dP\left( t\right) }{dt}=-\int_{0}^{t}\Pi \left( t-t'\right) P \left( t'\right) dt'+F\left( t\right) \text{,} \label {eq:GLE}
\end{equation}
where $P$ is the particle linear momentum, $\Pi \left( t\right) $ is the memory function, and $F\left( t\right) $ is a random force, which fulfills $\left\langle F\left( t\right) \right\rangle =0$ and the fluctuation-dissipation theorem (FDT)~\cite{Kubo66}:
\begin{equation}
\left\langle F\left( t\right) F\left( t'\right) \right\rangle=\left\langle P^{2}\right\rangle _{eq}\Pi \left( t-t'\right) \text{.} \label{eq:fdt}
\end{equation}
In most of the experimental situations which the EH does not holds arises in complex nonlinear or far from equilibrium structures where the detailed balance does not happens. Consequently, the FDT does not hold too. In supercooled liquids~\cite{Kauzmann48} and systems with activeted dynamics~\cite{Madrid03} we can find good examples. Those systems do not have apparently any easy analytical solution. Systems displaying glassy behavior have a multitude of metastable states that they visit after overcoming an irregular distribution of energy barriers, they exhibit aging and memory effects inherent to an activated dynamics characterized by a hierarchy of relaxation times~\cite{Madrid03}. On the another hand, diffusion can present closed solution for the main expectation values and arises as a simple laboratory for discuss those properties. In this case, as we shall see, Lapas \emph{et al.} give a full description of the KT validity~\cite{Lapas08}. In principle, it is generally possible to derive a GLE for Markovian systems by eliminating variables, whose effects are incorporated in the memory kernel and in the colored noise~\cite{Helleman80}. Altogether, some results obtained for the GLE formalism should be valid for diffusion described by fractional Fokker-Planck equations, since both formalisms yield similar results~\cite{Kou04,Dua08}.

\subsection{Khinchin theorem}
Since at the Khinchin times, most of the process studied were driven to the normal diffusion with exponential relaxation. Thus, it is quite natural to inquire if the KT will be affected, for instance in the slow relaxation dynamics that occurs in anomalous diffusion, where power laws, stretched exponential, Bessel functions and even a large jungle of functional behavior is possible for relaxation, besides the velocities distributions may not be a Gaussian.

Let us consider the Relaxation function
\begin{equation}
R(t) = \frac{C_P(t)}{C_P(0)}\text{,} \label{eq:R}
\end{equation}
where $C_P(t)$ is the correlation function for the dynamical operator $P$, which is defined as
\begin{equation}
C_P(t) = \left\langle P\left( t\right) P\left( 0\right)\right\rangle - \left\langle P\left( t\right) \right\rangle \left\langle P\left( 0\right) \right\rangle \text{.} \label{eq:CP}
\end{equation}
Explicitly the KT states that a system shall be ergodic in $P$ as long as
\begin{equation}
R(t \rightarrow \infty)=0\text{.} \label{eq:Ir}
 \end{equation}
In other words, irreversibility is a necessary and sufficient condition for EH. For macroscopic systems with a large number of degrees of freedom, the effect of past values of the forces usually vanishes for a sufficiently large $t$ and the aforementioned condition is quite reasonable.

Providing the well established method of recurrence relation~\cite{Lee83}, Lee has attempted to the fact that irreversibility is a broader concept than ergodicity and, in consequence, the KT may not work for all systems~ \cite{Lee07a}.  Physically, one can present some few systems where KT does not work, for example in the ordered phase of a magnetic system Eq. (\ref{eq:Ir}) is fulfilled, nevertheless the system never goes through over all states, since the symmetry breaking allows only specific states. In this sense, Lee's statement creates a new challenge namely ``in what systems the KT works?'' We shall show that the KT works for all range of anomalous diffusion described by a GLE, even in the presence of long range memory. The time averages of correlation functions are crucial for elucidating the properties of dynamical processes and play an extremely important role in the ergodic theory and, consequently, in physics. For diffusive systems governed by the GLE we shall see that the condition $R(t\rightarrow\infty)=0$ is sufficient for the time average to be equivalent to the ensemble average, \emph{i.e.}, for the system to be ergodic.

For any initial distribution of values, $P\left( 0\right) $ and $\left\langle F(t) (0) \right\rangle =0$, it is possible to obtain the temporal evolution of the moments of $P$ from GLE, Eq. (\ref{eq:GLE}),
\begin{equation}
\left\langle P\left( t\right) \right\rangle = \left\langle P\left( 0\right)\right\rangle R\left( t\right) \text{,} \label{eq:P}
\end{equation}
and
\begin{equation}
\left\langle P^{2}\left( t\right) \right\rangle =\left\langle P^{2}\right\rangle _{eq}+R^{2}\left( t\right) \left[ \left\langle P^{2}\left( 0\right) \right\rangle -\left\langle P^{2}\right\rangle _{eq} \right] \text{.} \label{eq:P2}
\end{equation}
To study relaxation we need to know $R(t)$, which can be calculated analytically in restricted cases, being obtained  numerically most of the times.  Consequently, one can describe completely those average values by the knowledge of $R(t)$. Higher order moments can be obtained, but is not our goal here. Since Eqs. (\ref{eq:P}) and (\ref{eq:P2}) are sufficient to show the validity of the KT for diffusion. If condition (\ref{eq:Ir}),
$R\left( t \rightarrow \infty \right)=0$, is valid, then the time evolution will produce the ensemble average with $\left\langle P\left( t\right) \right\rangle =0$ and $\left\langle P^{2}\left( t\right) \right\rangle = \left\langle P^{2}\right\rangle _{eq}$, \emph{i.e.} EH holds, then the KT is valid. Now we may ask in what situation the Eq. (\ref{eq:Ir}) is not valid. First, one should note that the long time behavior is associated with the small values of $z$ in the Laplace transform. Indeed, from the final value theorem we have
\begin{equation}
\lim_{t\rightarrow \infty }R\left( t\right) =\lim_{z\rightarrow 0}z\tilde{R}\left( z\right) \text{.} \label{final}
\end{equation}
Here we shall nominate the Laplace transform of a generic $G(t)$ by $\tilde{G}(z)$. Consequently it is only necessary to know $\tilde{R}(z)$. Now taking the equation of motion, Eq. (\ref {eq:GLE}), multipling by $p(0)$ and taking the ensemble average, we obtain
\begin{equation}
\frac{dR\left( t\right) }{dt}=-\int_{0}^{t}\Pi \left( t-t^{\prime}\right) R\left( t'\right) dt' ,\label {eq:dR}
\end{equation}
which Laplace transform yields $\tilde{R}\left( z\right) =1/[ z+\tilde{\Pi}\left( z\right) ] $. Morgado \emph{et al.}~\cite{Morgado02} have shown that if
\begin{equation}
\tilde{\Pi}\left( z\rightarrow0\right) \approx cz^{\nu }\text{,}\label{eq:Piz} 
\end{equation}
where $c$ is a positive constant, then $\alpha =1+\nu$. Consequently, the behavior of the memory $\tilde{\Pi}\left( z\right)$ for small $z$ determines the long range behavior of the diffusion, Eq. (\ref{eq:x2}).
Now the limit into Eq. (\ref{final}) reads
\begin{equation}
\lim_{t\rightarrow \infty }R\left( t\right) =\lim_{z\rightarrow 0} \left( 1+c\ z^{\alpha-2}\right)^{-1} \text{.} \label{limR2}
\end{equation}
For most of the diffusive process $0 < \alpha <2$, the irreversibility condition $R\left( t\rightarrow \infty \right) =0$ occurs. However, this condition fails for ballistic motion, $\alpha=2$, when $R( t \rightarrow \infty )=1/( 1+c)$ and the time-correlation function will be non-null for long times. Thus, if the ballistic system is not initially equilibrated, then it will never reach equilibrium and the final result of any measurement will depend on the initial conditions. In this situation the EH will not happens, however once again the KT works since the violation of the EH was induced by the violation of the irreversible condition, Eq. (\ref{eq:Ir}), as predicted by Khinchin. In other words, the ballistic diffusion violates ergodicity and the FDT~\cite{Costa03}. The major consequence of the violation of condition, Eq. (\ref{eq:Ir}), is the presence of a residual current. Let us suppose that the system starts with an average current such that $\left\langle P\left( 0\right) \right\rangle \neq 0$. From Eq.~(\ref{eq:P}), we obtain that for $R\left( t\rightarrow \infty \right) \neq 0$ a residual current remains. However, the effective current can be very small compared to $\left\langle P\left( 0\right) \right\rangle$ and its value, as any other measurable property for ballistic diffusion, will depend on the value of $c$. In other words, the system decays to a metastable state and remains in it indefinitely, even in the absence of an external field. For $\alpha >2$, $R( t \rightarrow \infty )=1$ , no memory of the initial conditions is lost, the process is not a diffusive one, and it is an activated process for which the GLE does not work \cite{Costa03}. Therefore, the  results of this letter apply to all kinds of diffusion, $0 < \alpha \leq 2$, described by a GLE independent of the memory range, which gives origin to new studies in ballistic diffusion~\cite{Bao03,Bao06,Lapas07}.

For stationary systems, $\chi(t,t')=\chi(t-t')$, one can define the time average integral as
\begin{equation}
I_{ta}=\lim_{T\rightarrow\infty}\frac{1}{T}\int_{0}^{T}\int_{0}^{t}\chi(t,t')dt'dt \text{.} \label{Ita}
\end{equation}
 Integrating by parts, we arrive at~\cite{Lee07a,Lee01}
\begin{equation}
I_{ta}=\lim_{t\rightarrow\infty}\left[\int_{0}^{t}\chi(t')dt'+R(t)-\frac{1}{t}\int_{0}^{t}R(t')dt'\right].\label{Ita2}
\end{equation}
Given that $R(t)$ is a real-valued function that converges asymptotically to a finite value, since we are dealing with the linear momentum autocorrelation, we can use a generalization of the final-value theorem for Laplace transforms~\cite{Gluskin03},
\[
\lim_{z\rightarrow0}z\tilde{R}(z)=\lim_{T\rightarrow\infty}\frac{1}{T}\int_{0}^{T}R(t)dt.
\]
From the above relation, we obtain from Eq.~(\ref{Ita2})
\begin{equation}
\tilde{\chi}(0)+R(t\rightarrow\infty)-\lim_{z\rightarrow0}z\tilde{R}(z)=\chi_{s}\text{,}\label{tildeChi1}
\end{equation}
where $\chi_{s}$ is the time independent value, the so-called static susceptibility. On the other hand, taking the Laplace transform of Eq.~(\ref{eq:dR}), we obtain $ \tilde{\chi}(z)+z\tilde{R}(z)=\chi_{s}$. Taking the limit $z\rightarrow0$, the previous relation becomes
\begin{equation}
\tilde{\chi}(0)+\lim_{z\rightarrow0}z\tilde{R}(z)=\chi_{s}\text{.}\label{tildeChi2}
\end{equation}
Comparing Eq.~(\ref{tildeChi2}) with Eq.~(\ref{tildeChi1}), one should conclude that the EH can only be valid if $R(t\rightarrow\infty)=0$, \emph{i.e.} if the irreversibility condition (\ref{eq:Ir}) holds. From Eq.~(\ref{final}) we end up with
\begin{equation}
\tilde{\chi}(0) = \chi_{s}.
\end{equation}
Again this is a consequence of the irreversibility condition. Therefore, irreversibility is a necessary and sufficient condition for the EH to hold in diffusive processes described by a GLE.

For nonlinear Hamiltonian maps~\cite{Shlesinger93} there is no general framework to address the problem and, in particular, the absence of a  coupling to a thermal bath (explicit in the GLE) and consequently the lack of a detailed balance relation or FDT may require a specific analysis of each case. However, since it is possible to give a kinetic description of the Hamiltonian dynamics by means of a fractional Fokker-Planck-Kolmogorov equation~\cite{Zaslavsky02}, it is expected that the treatment of anomalous diffusion in such systems should also be possible by the GLE formalism. Further research in this direction is needed and will open new perspectives.

\subsection{Nonexponential behavior}
From the above results, it is quite clear that the correlation for ballistic diffusion will not decay exponentially to equilibrium. Besides the ballistic case, any anomalous regime will present nonexponential decay. Even for normal diffusion, a large number of relaxations may be nonexponential~\cite{Morgado02}. There are a large number of phenomena where the systems do not relax immediately to equilibrium. Those phenomena, usually  associated with non-aging, have nonexponential relaxation and are most commonly described by power  laws or stretched exponentials.  The study of anomalous relaxation has produced  quite interesting results~\cite{Kauzmann48,Ricci00,Metzler00,Metzler04,Holek01,Rubi03,Hentschel07}.

For a system described by a GLE of the form Eq.~(\ref{eq:GLE}), the evolution relies on the noise that drives the particles. For a harmonic noise~\cite{Morgado02}
\begin{equation}
F(t)=\frac{1}{\sqrt{2k_BT}}\int  \sqrt{\rho(\omega)}\cos\left[ \omega t+\phi(\omega)\right] d\omega,
\label{noise}
\end{equation}
where $\rho$ is the noise spectral density, $k_B$ is the Boltzmann constant, and $\phi$ is a set of random phases in the range  $0\leq \phi \leq 2\pi$.

A systematic study carried on by Vainstein \emph{et al.}~\cite{Vainstein06} has shown that the spectral density plays a fundamental role in the description of stochastic processes as we shall see. First, the memory function $\Pi(t)$ can be easily obtained by the use of the FDT, Eq.~(\ref{eq:fdt}), as
\begin{equation}
\Pi(t) = \int  \rho(\omega)\cos(\omega t)d\omega\text{,}\label{eq:Pi}
\end{equation}
in such a way that the average cancels the random terms and obviously the memory is a deterministic even function. The Laplace transform  of the memory $\tilde{\Pi}(z)$ is an odd function in $z$; therefore,  $\tilde{R}(z)$ is also an odd function in $z$. This implies that by inverting the Laplace transform, $R(t)$ is an even function of $t$
\begin{equation}
R(-t) = R(t) \label{even}.
\end{equation}

Second, the condition given by Eq.~(\ref{eq:Piz}) to determine the exponent $\alpha$ ($\alpha=1+\nu$), for a spectral density of the form
\begin{equation}
\rho(\omega) = \left\{
\begin{array}{cc}
a \omega^{\beta}, &\text{ for } \omega\leq\omega_{s}\\
g(\omega), & \text{ otherwise.}
\end{array}\right.
\end{equation}
Here, the function $g(\omega)$ is arbitrary as long as it is sufficiently well-behaved and that its integral in the memory function converges. If one is interested only in the long time behavior $t\gg 1/\omega_s$, it can be taken to be $0$. With this noise density of states, it is possible to simulate many diffusive regimes~\cite{Vainstein06a}. Noise of this form can be obtained either by formal methods or empirical data. Using this expression in Eq.~(\ref{eq:Pi}), taking the Laplace transform in the limit $z \rightarrow 0$, we have 
\begin{equation}
\tilde{\Pi}(z) \propto \left\{
\begin{array}{cc}
z^{\beta}, &\text{ for } \beta<1,\\
-az\ln(z), &\text{ for } \beta=1,\\
z, & \text{ for } \beta>1.
\end{array}\right.
\end{equation}
Consequently, for this type of noise, there is a maximum value of $\alpha$, \emph{i.e.}, $\alpha\leq 2$ for any value of $\beta$. It should be noted that the case $\beta=1$ does not lead to a memory whose Laplace transform is in the form of Eq.~(\ref{eq:Piz}).  For $-1<\beta<1$, we obtain $\alpha = 1 + \beta$. For $\beta>1$, one has $\alpha = 2$, which shows that ballistic diffusion is a limiting case for the GLE with this type of memory. In this way, we can emphasize that
\begin{equation}
\nu =
\left \{
  \begin{array}{ll}
   \beta, & \mbox{  $\beta < 1$};\\
   \\
    1, & \mbox{  $\beta \geq 1$}. \label{beta}
  \end{array}
 \right.
 \end{equation}
This shows that a noise spectral density in the form of a power law can produce only diffusive motion in the region $0< \alpha \leq 2$. Diffusive motion beyond ballistic is not allowed, what can be observed by using the Laplace transform.

A recent study shows that exponential decay, power laws or even Mittag-Leffler functions are particular cases of  a more general function~\cite{Vainstein06} which approximates the decay. Indeed, considering time reversal symmetry~\cite{Oliveira81} and the discussion above (see Eq.~(\ref{even})), the correlation function must be even and cannot be any of those forms. We shall expose here the conditions under which it is possible to obtain an approximately exponential decay, what happens in certain circumstances for normal diffusion. In this case, 
\begin{equation}
 \gamma =  \lim_{z \rightarrow 0}\tilde{\Pi}(z) = \frac{\pi}{2}\rho(0).
\end{equation}
Consequently, the friction in the usual Langevin equation is nothing more than the noise spectral density for the lower modes. For an arbitrary memory, the system has a rich behavior; even for normal diffusion it is possible to show the existence of at least three time ranges~\cite{Vainstein06}.  A normal diffusion can be obtained using a spectral density of the form~\cite{Morgado02}
\begin{equation}
\rho(\omega) =
\left \{
  \begin{array}{ll}
   \frac{2\gamma}{\pi}, & \mbox{  $\omega < \omega_{s}$};\\
    0, & \mbox{ $\omega > \omega_{s}$} . \label{beta1}
  \end{array}
 \right.
 \end{equation}
For a broad band noise spectral density, $\gamma / \omega_{s} < 1$, and long times $t > \gamma^{-1}$ it is possible to decouple Eq.~(\ref{eq:dR}) to obtain an exponential decay of the form  $R(t)=\exp{(-\gamma t)}$, which will bring the system to equilibrium, that is $R(t\rightarrow\infty ) \rightarrow 0$.

\subsection{Evidencing the KT}
In order to illustrate the analytical results, we have numerically integrated the GLE, Eq.~(\ref{eq:GLE}), to obtain approximations to the probability distribution of particle velocities using histograms.
\begin{figure}[h!]
\includegraphics[scale=0.45]{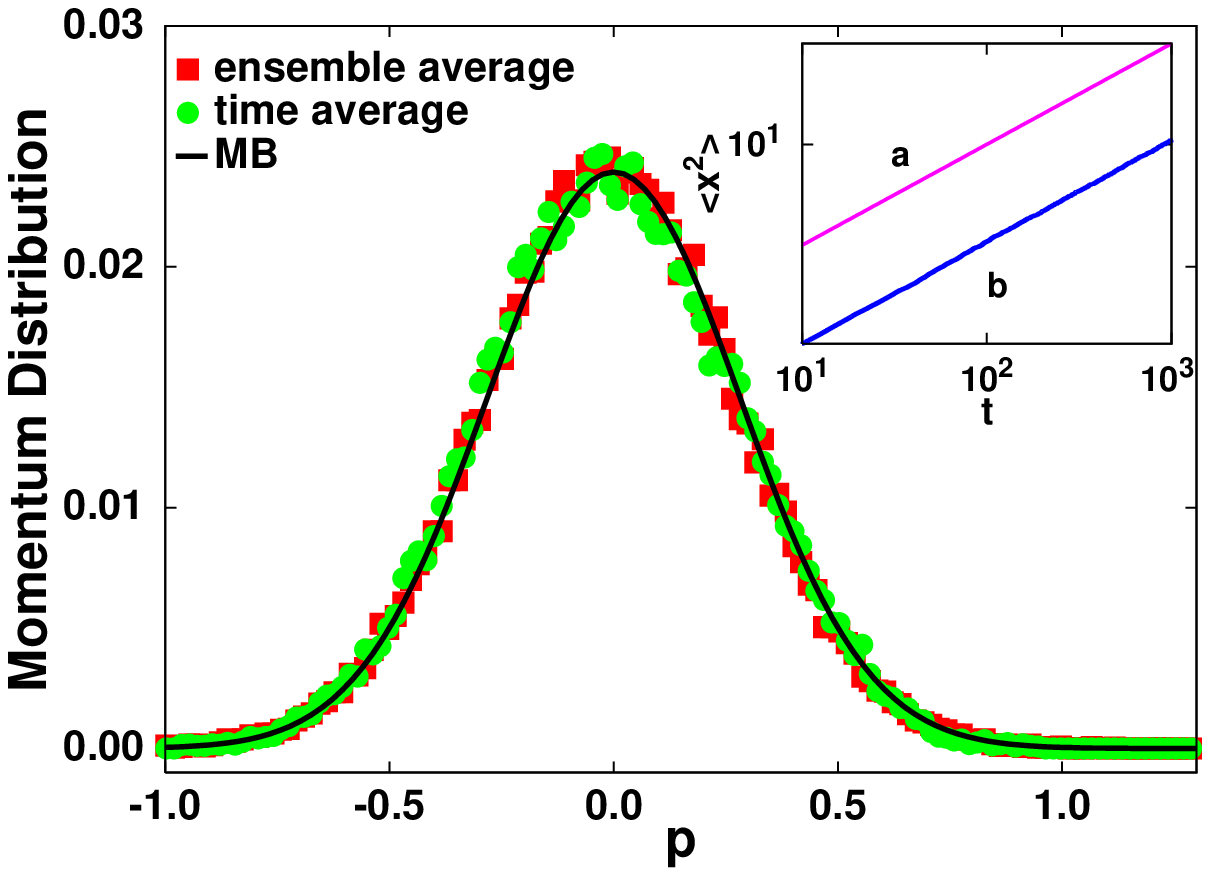}
\includegraphics[scale=0.45]{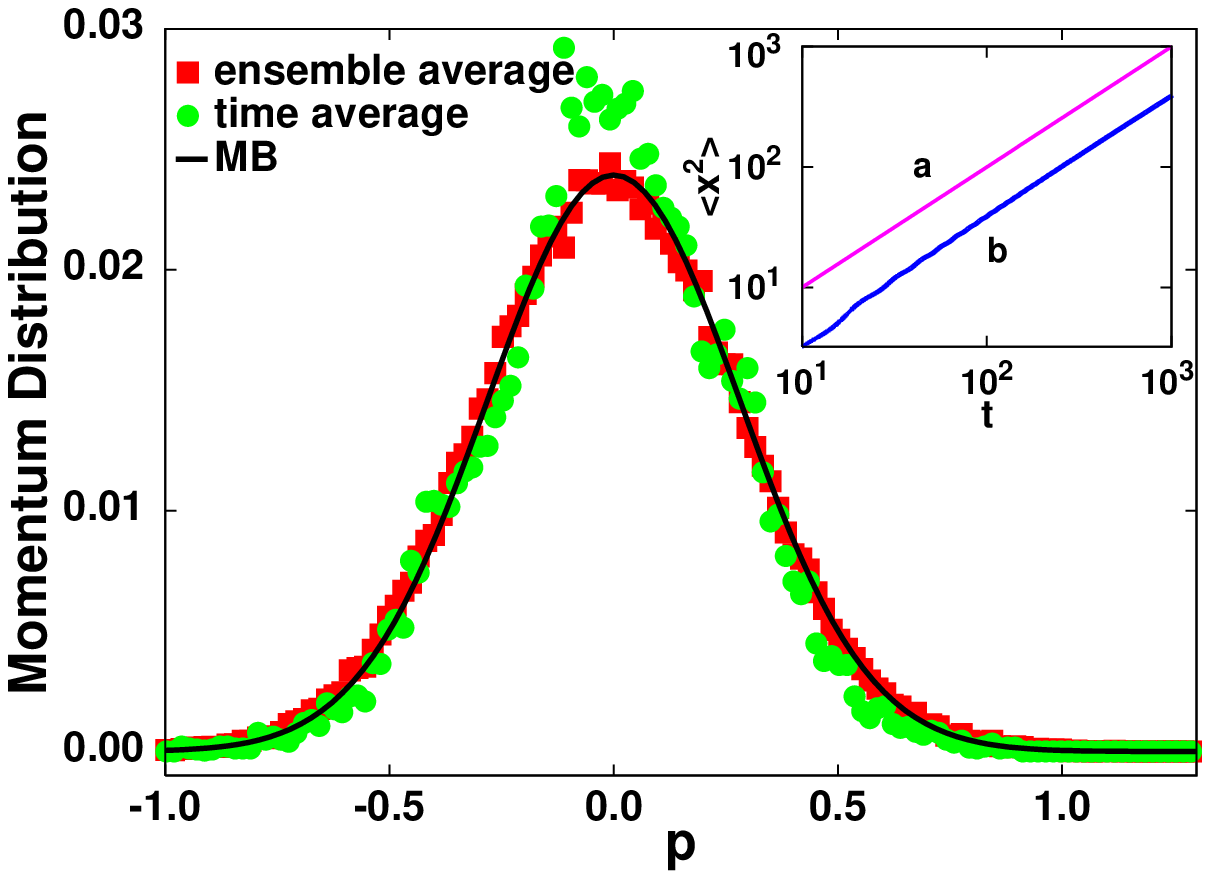}
\includegraphics[scale=0.45]{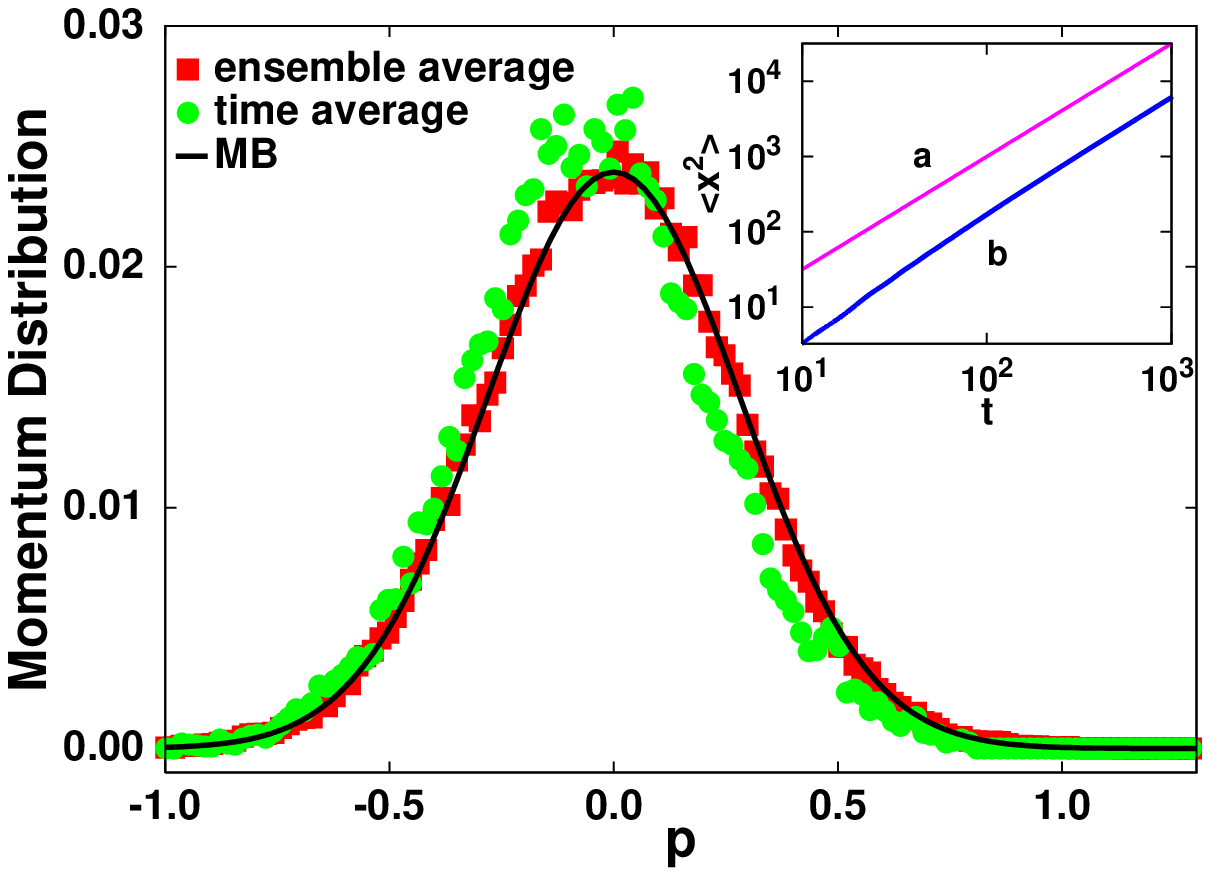}
\caption{(Color online) Numerical results for the probability distribution function for subdiffusion (top, $\alpha=0.5$), normal diffusion (middle, $\alpha=1$) and superdiffusion (bottom, $\alpha=1.5$). The time averages (circles) are obtained by following one particle trajectory and calculating the histogram for times from $t=100$ to $t=5000$. For the ensemble averages (squares), we calculate the histogram using $5\cdot10^{4}$ particles, at time $t=1000$. The continuous line is the Maxwell-Boltzmann distribution. Insets: Curves $a$ correspond to the functions $t^\alpha$ and curves $b$ to the simulated mean square displacements.}
\label{figure}
\end{figure}

In Fig.~(\ref{figure}) we show the probability distribution functions obtained for subdiffusion ($\beta=-0.5$), normal diffusion ($\beta=0$), and superdiffusion ($\beta=0.5$) for the values $a=0.25$ and $g(w)=0$. We have used $\omega_s=0.5$ for all cases except for subdiffusion, which demands a broad noise $\omega_s=2$ to reach the stationary state. In all cases, we expect that $R(t\rightarrow\infty)=0$, and that the EH will be valid even for the subdiffusive (superdiffusive) case, despite the fact that $W=0$ ($W=\infty$). These relations can be seen by considering the limit $W = \lim_{z\rightarrow 0} \tilde{R}(z)$. If the EH is valid, the velocity probability distribution will be the same for an average over an ensemble of particles and for a time average over the trajectory of a single particle for long times after the system has reached an equilibrium state. Note that despite the presence of large fluctuations in the time average case due to numerical errors, there is a good agreement between the resulting ensemble and time distributions. The three probability distributions converge toward the Maxwell-Boltzmann distribution,  which is in accordance with previous analytical results~\cite{Lapas07}.

\section{Remarks}
In order to understand cellular life processes like cell division and nerve cell growth, it has been shown that the protein reaches the position via a random, one-dimensional diffusion movement along the microtubule~\cite{Helenius06}. Besides that, statistical analysis leads to purely diffusive dynamic laws such that some internal processes induce a breakdown of the central-limit theorem and leads to anomalous diffusion~\cite{Faretta98}. Those molecular motors are trapped in a rough energy landscape subjected to a kind of glassy behavior common to a great variety of complex systems such as glasses~\cite{Rubi97} and proteins~\cite{Frauenfelder91,Onuchic97}, just to mention some examples, which corresponds to apply an external force strong enough to place the system near the stall point of the motor~\cite{Kafri05}.

A large number of theorems in Statistical Physics and even a proper applications of formalism such as a "simple linear response theory" relay under the ergodic hypothesis and its validation. The Khinchin theorem put the EH in a straight way, and it has a practical character, since it is exposed in therm of response function. In this way, the Lee's works need a large discussion concerning the validity of the KT in specific systems. In this work we have discussed the KT for anomalous diffusion, which are ergodic in the range of exponents $0 \leq \alpha \leq 2$, where $\alpha$ defines the asymptotic behavior of the diffusion, Eq. (\ref{eq:x2}). For $\alpha=2$ we have the special ballistic case as we have seen the EH does not work for ballistic diffusion, however, the condition Eq. (\ref{eq:Ir}) does not work either, consequently the KT remains as a strong reference, at least for diffusion. In recent years Molecular motors has been driving a lot of attention~\cite{Bao03,Bao06} with a large potential for pure and applied science, there those discussion may be very useful. There are many situations of violation of the EH, particularly in glassy systems \cite{Ricci00}, and another were EH holds, for example recently~\cite{Hentschel07} dynamical simulations and equilibrium statistical mechanics were used to treat glass transition calculations. The agreement between them is a strong indication of the validity of the HE. This is a quite surprising result for such complex system. Disordered systems are a large universe to explore the basic assumptions of statistical mechanics in particular of the KT.

\acknowledgments
 This work was supported by FAPDF, CAPES, and CNPq.

\bibliographystyle{unsrt}

\end{document}